\documentclass[12pt, draftclsnofoot, onecolumn]{IEEEtran}
\usepackage{epsfig,graphicx,subfigure,psfrag,amsmath,cases,bm}
\usepackage{latexsym,amssymb,algorithm,mathtools}
\usepackage{algorithmic}
\usepackage{color}
\usepackage{url}
\usepackage{cite}
\usepackage{scrtime}
\usepackage{stfloats}
\usepackage{tablefootnote}
\usepackage{geometry}
\allowdisplaybreaks

\author{\IEEEauthorblockN {Dongfang Xu\IEEEauthorrefmark {1}, Xianghao Yu\IEEEauthorrefmark {4}, Derrick Wing Kwan Ng\IEEEauthorrefmark {3}, and Robert Schober\IEEEauthorrefmark {1}}

\IEEEauthorrefmark {1}Friedrich-Alexander-University
Erlangen-N\"urnberg, Germany\\
\IEEEauthorrefmark {4}Dept. of ECE, The Hong Kong University of Science and Technology, Hong Kong\\
\IEEEauthorrefmark {3}The University
of New South Wales, Australia

}

\newtheorem{T-Prob}{Transformed Problem}

\DeclareMathOperator{\mino}{minimize}

\newcommand{\qed}{\hfill \ensuremath{\blacksquare}}
\newtheorem{Remark}{Remark}

\title{Resource Allocation for Active IRS-Assisted Multiuser Communication Systems}

\begin{document}
\maketitle
\begin{abstract}
Intelligent reflecting surfaces (IRSs) are emerging as promising enablers for the next generation of wireless communication systems, because of their ability to customize favorable radio propagation environments. However, with the conventional passive architecture, IRSs can only adjust the phase of the incident signals limiting the achievable beamforming gain. To fully unleash the potential of IRSs, in this paper, we consider a more general IRS architecture, i.e., active IRSs, which can adapt the phase and amplify the magnitude of the reflected incident signal simultaneously with the support of an additional power source. To realize green communication in active IRS-assisted multiuser systems, we jointly optimize the reflection matrix at the IRS and the beamforming vector at the base station (BS) for the minimization of the BS transmit power. The resource allocation algorithm design is formulated as an optimization problem taking into account the maximum power budget of the active IRS and the quality-of-service (QoS) requirements of the users. To handle the non-convex design problem, we develop a novel and computationally efficient algorithm based on the bilinear transformation and inner approximation methods. The proposed algorithm is guaranteed to converge to a locally optimal solution of the considered problem. Simulation results illustrate the effectiveness of the proposed scheme compared to two baseline schemes. Moreover, the results unveil that deploying active IRSs is a promising approach to enhance the system performance compared to conventional passive IRSs, especially when strong direct links exist.
\end{abstract}
\section{Introduction}
Future wireless communication systems are envisioned to provide high data-rate communication services \cite{wong2017key}. Inspired by recent advances in electromagnetic metamaterials, revolutionary new metasurfaces, called intelligent reflecting surfaces (IRSs) have been proposed for deployment in conventional communication networks to satisfy this demand \cite{cui2014coding}. In particular, comprising a number of programmable elements, IRSs can be smartly adapted to the channel conditions so as to proactively customize the radio propagation environment for enhancing the system performance \cite{yu2021smart}. Moreover, due to the passive nature of the reflecting elements, e.g., diodes and phase shifters, the power required for maintaining the IRS operation is typically very small \cite{cui2014coding}. Furthermore, commonly fabricated as thin rectangular surfaces, IRSs can be flexibly deployed coexisting with existing infrastructure and smoothly integrate into conventional communication systems. 
\par
These favorable properties have motivated numerous works to study IRSs for performance enhancement of conventional communication systems \cite{wu2019intelligent,8741198,yu2020power}. Yet, in practice, the end-to-end path loss of the BS-IRS-receiver link is in general much larger than that of the unobstructed direct link due to the double path loss effect \cite{9306896}. Hence, employing passive IRSs may not effectively enhance the system performance. To compensate for the severe double path loss in the cascaded IRS channel, one has to adopt a large passive IRS comprising hundreds if not thousands of phase shift elements to achieve a significant passive beamforming gain \cite{9306896}, \cite{xu2021optimal}. However, deploying a large number of passive IRS elements significantly increases the signaling overhead for channel estimation and the complexity of IRS optimization \cite{wu2019intelligent,8741198,yu2020power}, which makes the design of IRS-assisted wireless systems challenging in practice. To circumvent these issues, the authors of \cite{zhang2021active} recently proposed a new IRS structure, namely, active IRSs. In particular, equipped with reflection-type amplifiers \cite{lonvcar2019ultrathin}, \cite{you2021wireless}, active IRSs can not only reflect the incident signals by manipulating the programmable IRS elements, but also amplify the reflected signal with the support of an extra power supply. We note that active IRSs are fundamentally different from full-duplex amplify-and-forward (FD-AF) relays in terms of hardware architecture and the mode of transmission. Specifically, equipped with radio frequency (RF) chains, FD-AF relays are able to receive the incident signal and then transmit it after amplification at the expense of self-interference. This process introduces a delay incurred by the signal processing at the relay. In contrast, equipped with low-power reflection-type amplifiers, active IRSs reflect and amplify the incident signal instantaneously, and the resulting delay between the direct link and the reflected link is negligibly small compared to the symbol duration \cite{wu2019intelligent}. Moreover, the signals received at different relay antennas are jointly amplified via an amplification matrix. In contrast, for active IRSs, the signals received at different IRS elements are individually amplified. On the other hand, compared to conventional passive IRSs, active IRSs can effectively compensate the double path loss effect without significantly complicating the IRS design \cite{zhang2021active}. To illustrate this, the authors of \cite{zhang2021active} studied the joint transmit and reflect beamforming design for maximization of the spectral efficiency of an active IRS-assisted multiuser communication system. The resource allocation algorithm design was formulated as a series of quadratic constraint quadratic programming (QCQP) problems which were tackled in an alternating manner. In fact, to realize the potential gains facilitated by active IRSs, the appropriate amount of power has to be assigned to each element of the active IRS from the limited available power. As a result, compared to systems assisted by conventional passive IRSs, it is more important to delicately design the BS beamforming such that the power consumption of the whole system is still affordable and the quality-of-service (QoS) requirements of the users can be satisfied. Alternating optimization (AO)-based optimization frameworks cannot effectively handle the aforementioned power minimization problem. In particular, such problems cannot be easily transformed to standard QCQP or second-order cone program (SOCP) problems with convex constraints that can be efficiently solved by employing AO-based algorithms \cite{wu2019intelligent}, \cite{zhang2021active}. Moreover, by dividing the coupled optimization variables into disjoint groups, AO-based algorithms inevitably eliminate the joint optimality of the BS beamforming vectors, the IRS amplification factor matrix, and the IRS phase shift matrix in the considered power minimization problem, which may lead to unsatisfactory performance \cite{bezdek2002some}. Besides, for the considered power minimization problem, the monotonicity of the objective value during AO cannot be guaranteed because of the required Gaussian randomization \cite{yu2020power}.
\par
Motivated by the above discussion, in this paper, we investigate the resource allocation algorithm design for active IRS-assisted communication systems, where the active IRS can amplify the reflected signal exploiting an additional power source. To this end, we aim to minimize the transmit power of the BS by jointly designing the BS beamformers and the IRS reflection matrix, taking into account the QoS requirements of the users and the maximum power budget of the active IRS. Since the optimization variables are highly coupled in the resulting non-convex optimization problem, the corresponding globally optimal solution is challenging to obtain. As a compromise, by capitalizing on bilinear transformation, inner approximation, and semidefinite relaxation, we develop a novel iterative algorithm, which enjoys low computational complexity. The proposed algorithm is guaranteed to converge to a locally optimal solution of the considered problem. Our simulation results reveal that active IRSs are a promising solution to fully exploit the potential of IRS-assisted wireless systems, especially when non-negligible direct links exist.
\par
\textit{Notation:} Vectors and matrices are denoted by boldface lower case and boldface capital letters, respectively. $\mathbb{R}_+^{N\times M}$ and $\mathbb{C}^{N\times M}$ denote the spaces of $N\times M$ positive real-valued matrices and complex-valued matrices, respectively. $\Re\left \{ \cdot \right \}$ extracts the real part of a complex number. $|\cdot|$ and $||\cdot||$ denote the absolute value of a complex scalar and the Euclidean norm of its argument, respectively. $\mathbf{I}_{N}$ refers to the identity matrix of dimension $N$. $\mathbb{H}^{N}$ denotes the set of complex Hermitian matrices of dimension $N$. $\mathbf{A}^H$ refers to the conjugate transpose of matrix $\mathbf{A}$. $\mathbf{A}\succeq\mathbf{0}$ indicates that $\mathbf{A}$ is a positive semidefinite matrix. $||\mathbf{A}||_F$, $\mathrm{Tr}(\mathbf{A})$, and $\mathrm{Rank}(\mathbf{A})$ denote the Frobenius norm, the trace, and the rank of matrix $\mathbf{A}$, respectively. $\mathrm{diag}(\mathbf{a})$ represents a diagonal matrix whose main diagonal elements are extracted from vector $\mathbf{a}$; $\mathrm{Diag}(\mathbf{A})$ denotes a vector whose
elements are extracted from the main diagonal elements of matrix $\mathbf{A}$. $\mathcal{E}\left \{ \cdot \right \}$ represents statistical expectation. $\overset{\Delta }{=}$ and $\sim$ refer to ``defined as'' and ``distributed as'', respectively. $\mathcal{CN}(\mu ,\sigma^2)$ indicates the distribution of a circularly symmetric complex Gaussian random variable with mean $\mu$ and variance $\sigma^2$. $\mathbf{X}^*$ refers to the optimal value of optimization variable $\mathbf{X}$.
\section{System Model}
\begin{figure}[t]
\vspace*{-6mm}
\centering
\includegraphics[width=2.2in]{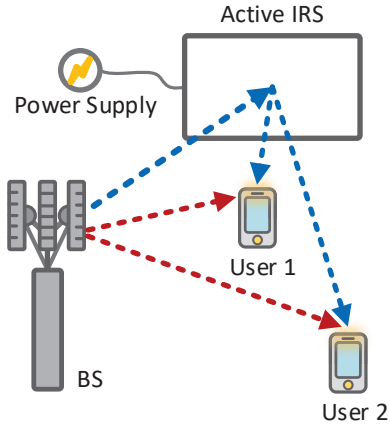} \vspace*{-6mm}
\caption{An active IRS-assisted communication system consist of one multi-antenna BS and $K=2$ users. The active IRS is supported by a power supply. The direct links and reflected links between the BS and the users are denoted by red dashed lines and blue dash lines, respectively.}
\label{system_model}\vspace*{-4mm}
\end{figure}
We consider an active IRS-assisted multiuser multiple-input single-output (MISO) communication system, cf. Figure \ref{system_model}. The BS is equipped with $N_\mathrm{T}$ antennas while all $K$ users are single-antenna devices. To enhance the performance of the considered system, an active IRS is employed to assist the information transmission from the BS to the users. In particular, the active IRS is composed of $M$ elements and is supported by an additional power source. Equipped with an integrated active reflection-type amplifier, each IRS element can not only smartly alter the phase of the incident signals, but also amplify the reflected signal for effective beamforming. To establish a performance upper bound for the considered system, we assume that the perfect channel state information (CSI) of the entire system is available at the BS. The CSI can be acquired with one of the existing channel estimation schemes proposed for IRS-assisted wireless systems \cite{9366805}, \cite{9087848}. To simplify the notation, we collect the indices of the users and IRS elements in sets $\mathcal{K}=\left \{1,\cdots ,K \right \}$ and $\mathcal{M}=\left \{1,\cdots ,M \right \}$, respectively.
\par
In each scheduled time slot, the signal vector $\mathbf{x}$ transmitted by the BS is constructed as follows
\begin{equation}
\mathbf{x}=\underset{k\in\mathcal{K}}{\sum }\mathbf{w}_kb_k,
\end{equation} 
where $\mathbf{w}_k\in\mathbb{C}^{N_\mathrm{T}\times 1}$ and $b_k\in \mathbb{C}$ denote the beamforming vector for user $k$ and the corresponding information symbol. We assume $\mathcal{E}\{\left |b_k\right|^2\}=1$, $\forall\mathit{k} \in \mathcal{K}$, without loss of generality.
\par
Employing reflection-type amplifiers \cite{lonvcar2019ultrathin} driven by a common power supply, the signal reflected and amplified by the active IRS is given by
\begin{equation}
\mathbf{y}=\mathbf{A}\bm{\Theta}\mathbf{G}\mathbf{x}+\underbrace{\mathbf{A}\bm{\Theta}\mathbf{d}}_{\text{dynamic noise}}+\underbrace{\mathbf{s}}_{\text{static noise}},\label{IRSsignal}
\end{equation}
where $\mathbf{A}\overset{\Delta}{=}\mathrm{diag}(a_1,\cdots,a_M)\in\mathbb{R}^{M\times M}_+$ and $\bm{\Theta}\overset{\Delta}{=}\mathrm{diag}(e^{j\psi_1 },\cdots,e^{j\psi _M})\in\mathbb{C}^{M\times M}$ denote the amplification factor matrix and the phase shift matrix of the active IRS, respectively. Matrix $\mathbf{G}\in\mathbb{C}^{M\times N_\mathrm{T}}$ denotes the channel between the BS and the IRS. Moreover, we observe from \eqref{IRSsignal} that the noises at the IRS can be divided into two categories, i.e., dynamic noise and static noise \cite{zhang2021active}. In particular, the dynamic noise is generated due to the power amplification \cite{you2021wireless}, where $\mathbf{d}\in\mathbb{C}^{N_\mathrm{T}\times 1}$ is modelled as additive white Gaussian noise (AWGN) with variance $\sigma_d^2$, i.e., $\mathbf{d}\sim\mathcal{CN}(\mathbf{0}_{N_\mathrm{T}},\sigma_d^2\mathbf{I}_{N_\mathrm{T}})$ \cite{zhang2021active}. The static noise $\mathbf{s}\in\mathbb{C}^{N_\mathrm{T}\times 1}$ is modelled as AWGN with variance $\sigma_s^2$, i.e., $\mathbf{s}\sim\mathcal{CN}(\mathbf{0}_{N_\mathrm{T}},\sigma_s^2\mathbf{I}_{N_\mathrm{T}})$, and it is not affected by $\mathbf{A}$ and its power is usually negligibly small compared to that of the dynamic noise $\mathbf{A}\bm{\Theta}\mathbf{d}$ \cite{6047578}.
\par
The received signal at user $k$ is given by
\begin{eqnarray}
r_k\hspace*{-6mm}&&=\underbrace{(\mathbf{h}_{\mathrm{D},k}^H+\mathbf{h}_{\mathrm{R},k}^H\mathbf{A}\bm{\Theta}\mathbf{G})\mathbf{w}_kb_k}_{\text{desired signal}}+\underbrace{(\mathbf{h}_{\mathrm{D},k}^H+\mathbf{h}_{\mathrm{R},k}^H\mathbf{A}\bm{\Theta}\mathbf{G})\underset{\substack{r\in\mathcal{K}\\r\neq k}}{\sum }\mathbf{w}_rb_r}_{\text{multiuser interference}}\notag\\
\hspace*{-6mm}&&+\underbrace{\mathbf{h}_{\mathrm{R},k}^H\mathbf{A}\bm{\Theta}\mathbf{d}}_{\text{dynamic noise introduced by the IRS}}+\underbrace{n_k}_{\text{noise introduced at user $k$}},
\end{eqnarray}
where $\mathbf{h}_{\mathrm{D},k}\in\mathbb{C}^{N_\mathrm{T}\times 1}$ and $\mathbf{h}_{\mathrm{R},k}\in\mathbb{C}^{M\times 1}$ denote the channel vectors of the BS-user $k$ link (direct link) and the IRS-user $k$ link (reflected link), respectively. $n_k$
represents the AWGN at the user $k$ with zero mean and variance $\sigma_{n_k}^2$, i.e., $n_k\sim\mathcal{CN}(0,\sigma_{n_k}^2)$.
\par

\section{Problem Formulation}
The received signal-to-interference-plus-noise ratio (SINR) of user $k$ is given by
\begin{eqnarray}
\Gamma _k=
    \frac{\left | (\mathbf{h}_{\mathrm{D},k}^H+\mathbf{h}_{\mathrm{R},k}^H\mathbf{A}\bm{\Theta}\mathbf{G})\mathbf{w}_k\right |^2}{\underset{\substack{r\in\mathcal{K}\\r\neq k}}{\sum }\left | (\mathbf{h}_{\mathrm{D},k}^H+\mathbf{h}_{\mathrm{R},k}^H\mathbf{A}\bm{\Theta}\mathbf{G})\mathbf{w}_r\right |^2\hspace*{-1mm}+\sigma_d^2\left \|\mathbf{h}_{\mathrm{R},k}^H\mathbf{A}\bm{\Theta} \right \|^2\hspace*{-1mm}+\sigma_{n_k}^2}.
\end{eqnarray}
\par
In this paper, we aim to minimize the BS transmit power while satisfying the QoS requirements of the users and the maximum power allowance of the active IRS. In particular, the joint design of the BS beamforming vectors, the IRS amplification factor matrix, and the IRS phase shift matrix, i.e., $\left \{\mathbf{w}_k,\mathbf{A},\bm{\Theta} \right \}$, is obtained by solving the following optimization problem
\begin{eqnarray}
\label{prob1}
&&\hspace*{-12mm}\underset{\mathbf{w}_k,\mathbf{A},\bm{\Theta}}{\mino} \,\, \,\, \underset{k\in\mathcal{K}}{\sum }\left \|\mathbf{w}_k\right \|^2\notag\\
&&\hspace*{-12mm}\mbox{subject to}\hspace*{2mm}
\mbox{C1:}\hspace*{1mm}\Gamma_{\mathrm{req}_k}\leq\Gamma_k,\hspace*{1mm}\forall k,\hspace*{2mm}\mbox{C2:}\hspace*{1mm}\underset{k\in\mathcal{K}}{\sum }\left \|\mathbf{A}\bm{\Theta}\mathbf{G} \mathbf{w}_k \right \|^2+\sigma_d^2\left \|\mathbf{A}\bm{\Theta}\right \|_F^2\leq P_{\mathrm{A}}.
\end{eqnarray}
Here, $\Gamma_{\mathrm{req}_k}$ in constraint C1 is the minimum required SINR of user $k$. Constraint C2 indicates that the amplification power of the active IRS should be less than or equal to the maximum power allowance $P_{\mathrm{A}}$. We note that the optimization problem in \eqref{prob1} is non-convex due to the coupled optimization variables and the fractional constraint C1. Next, by employing the bilinear transformation and IA, we develop a iterative low-complexity algorithm which is guaranteed to converge to a locally optimal solution of the problem in \eqref{prob1}.
\par
\begin{Remark}
Compared to passive IRS design, though active IRS design sidesteps the unit-modulus constraint, it also introduces the additional non-convex constraint C2 which aggravates the coupling between the optimization variables. In fact, for resource allocation design for IRS-assisted systems, the coupling between the optimization variables is an unavoidable obstacle. For passive IRSs, such obstacle is commonly tackled by employing AO-based algorithms \cite{zhang2021active}, \cite{8930608} or IA-based algorithms \cite{yu2020power}. However, employing AO-based algorithms destroys the joint optimality of the optimization variables, which may lead to unsatisfactory system performance. Moreover, it has been shown in \cite{yu2020power} that for power minimization problems, the commonly adopted AO-based algorithm with Gaussian randomization is not guaranteed to generate a monotonically decreasing sequence of the objective function values during the iterations. On the other hand, when directly applying IA, the matrix $\bm{\Theta}=\mathrm{diag}(e^{j\psi_1 },\cdots,e^{j\psi _M})$ at the IRS is first transformed into a vector $\mathbf{v}=[e^{j\psi_1 },\cdots,e^{j\psi _M}]^H$ \cite{8930608}. Then, a new optimization variable $\mathbf{V}$ is defined as $\mathbf{V}=\mathbf{v}\mathbf{v}^H$, which imposes three additional constraints on the considered optimization problem, i.e., $\mathbf{V}\succeq \mathbf{0}$, $\mathrm{Diag}(\mathbf{V})=\mathbf{1}$, and a non-convex constraint $\mathrm{Rank}(\mathbf{V})=1$. In the literature, the rank-one constraint is usually removed by employing SDR. However, by doing so, the rank of the obtained solution is in general larger than one \cite{luo2010semidefinite}. Alternatively, $\mathrm{Rank}(\mathbf{V})=1$ can be equivalently transformed into a difference of norm functions, and then be tackled by a penalty-based algorithm  \cite{xu2020resource}. However, since the penalty factor cannot be infinitely large in practice, such an approach can only guarantee a suboptimal solution. To circumvent these obstacles, in this paper, for active IRSs, we employ bilinear transformation and IA and develop a low-complexity iterative algorithm which is guaranteed to converge to a locally optimal solution of the optimization problem in \eqref{prob1} \cite{marks1978general}.
\end{Remark}

\section{Solution of the Optimization problem}
\subsection{Bilinear Transformation}
Note that matrices $\mathbf{A}$ and $\bm{\Theta}$ in \eqref{prob1} always appear in product form. Hence, we rewrite the product term $\mathbf{A}\bm{\Theta}$ as $\bm{\Psi}=\mathrm{diag}(a_1e^{j\psi_1 },\cdots,a_Me^{j\psi _M})\in\mathbb{C}^{M\times M}$. Then, the quadratic term $\sigma_d^2\left \| \mathbf{h}_{\mathrm{R},k}^H\mathbf{A}\bm{\Theta}\right \|^2$ in constraint C1 can be rewritten as follows
\begin{equation}
    \sigma_d^2\left \| \mathbf{h}_{\mathrm{R},k}^H\mathbf{A}\bm{\Theta}\right \|^2=\sigma_d^2\mathrm{Tr}(\bm{\Psi}^H\mathbf{H}_{\mathrm{R},k}\bm{\Psi} ),
\end{equation}
where $\mathbf{H}_{\mathrm{R},k}\in\mathbb{C}^{M\times M}$ is defined as $\mathbf{H}_{\mathrm{R},k}=\mathbf{h}_{\mathrm{R},k}\mathbf{h}_{\mathrm{R},k}^H$.
To facilitate the application of the IA algorithm, we define $\mathbf{W}_k=\mathbf{w}_k\mathbf{w}_k^H$, $\forall k$, and rewrite the quadratic term $\left | (\mathbf{h}_{\mathrm{D},k}^H+\mathbf{h}_{\mathrm{R},k}^H\bm{\Psi}\mathbf{G})\mathbf{w}_r\right |^2$ in constraint C1 as follows
\begin{eqnarray}
\label{vectornormtotrace}
&&\left | (\mathbf{h}_{\mathrm{D},k}^H+\mathbf{h}_{\mathrm{R},k}^H\bm{\Psi}\mathbf{G})\mathbf{w}_r\right |^2\notag\\
=\hspace*{-4mm}&&\mathbf{h}_{\mathrm{D},k}^H\mathbf{W}_r\mathbf{h}_{\mathrm{D},k}+\mathbf{h}_{\mathrm{R},k}^H\bm{\Psi}\mathbf{G}\mathbf{W}_r \mathbf{G}^H\bm{\Psi}^H\mathbf{h}_{\mathrm{R},k}
+2\Re\left \{\mathbf{h}_{\mathrm{D},k}^H\mathbf{W}_r \mathbf{G}^H\bm{\Psi}^H\mathbf{h}_{\mathrm{R},k} \right \}\notag\\
=\hspace*{-4mm}&&\mathrm{Tr}\left ( \begin{bmatrix}
  \mathbf{h}_{\mathrm{R},k} \\ 
  \mathbf{h}_{\mathrm{D},k}
\end{bmatrix}
\begin{bmatrix}
  \mathbf{h}_{\mathrm{R},k}^H & \mathbf{h}_{\mathrm{D},k}^H
\end{bmatrix}
\begin{bmatrix}
  \mathbf{0} & \bm{\Psi}\mathbf{G}\mathbf{W}_r^H \\ 
  \mathbf{W}_r \mathbf{G}^H\bm{\Psi}^H & \mathbf{0}
\end{bmatrix}\right )\notag\\
+\hspace*{-4mm}&&\mathrm{Tr}(\mathbf{H}_{\mathrm{D},k}\mathbf{W}_r)+\mathrm{Tr}(\bm{\Psi}\mathbf{G}\mathbf{W}_r \mathbf{G}^H\bm{\Psi}^H\mathbf{H}_{\mathrm{R},k}),
\end{eqnarray}
where $\mathbf{H}_{\mathrm{D},k}\in\mathbb{C}^{N_\mathrm{T}\times N_\mathrm{T}}$ is defined as $\mathbf{H}_{\mathrm{D},k}=\mathbf{h}_{\mathrm{D},k}\mathbf{h}_{\mathrm{D},k}^H$. Then, we recast the optimization problem in \eqref{prob1} equivalently as follows
\begin{eqnarray}
\label{prob2}
&&\underset{\bm{\Psi},\mathbf{W}_k\in\mathbb{H}^{N_{\mathrm{T}}}}{\mino} \,\, \,\, \hspace*{2mm}\underset{k\in\mathcal{K}}{\sum }\mathrm{Tr}(\mathbf{W}_k)\notag\\
&&\mbox{subject to}\hspace*{4mm}
\mbox{C1:}\hspace*{1mm}\Gamma _k\geq \Gamma_{\mathrm{req}_k},\hspace*{1mm}\forall k,\notag\\
&&\hspace*{21mm}\mbox{C2:}\hspace*{1mm}\underset{k\in\mathcal{K}}{\sum }\mathrm{Tr}(\bm{\Psi}\mathbf{G} \mathbf{W}_k\mathbf{G}^H\bm{\Psi}^H)+\sigma_d^2\mathrm{Tr}(\bm{\Psi}\bm{\Psi}^H)\leq P_{\mathrm{A}},\notag\\
&&\hspace*{21mm}\mbox{C3:}\hspace*{1mm}\mathbf{W}_k\succeq\mathbf{0},\hspace*{1mm}\forall k,\hspace*{10mm}\mbox{C4:}\hspace*{1mm}\mathrm{Rank}(\mathbf{W}_k)\leq 1,\hspace*{1mm}\forall k. 
\end{eqnarray}
We note that the coupling between $\mathbf{W}_k$ and $\bm{\Psi}$ in constraints C1 and C2 and the rank-one constraint C4 are obstacles to solving \eqref{prob2}. Next, we take the term $\mathrm{Tr}(\bm{\Psi}\mathbf{G}\mathbf{W}_r \mathbf{G}^H\bm{\Psi}^H\mathbf{H}_{\mathrm{R},k})$ as an example to illustrate how to construct a convex subset for the non-convex constraint C1. Note the fact that for arbitrary matrices $\mathbf{C}$ and $\mathbf{D}$ having the same dimensions, we have $\mathrm{Tr}(\mathbf{C}\mathbf{D})=\frac{1}{2}\left \|\mathbf{C}+\mathbf{D} \right \|_F^2-\frac{1}{2}\mathrm{Tr}(\mathbf{C}^H\mathbf{C})-\frac{1}{2}\mathrm{Tr}(\mathbf{D}^H\mathbf{D})$. Hence, we first rewrite the coupling term $\mathrm{Tr}(\bm{\Psi}\mathbf{G}\mathbf{W}_r \mathbf{G}^H\bm{\Psi}^H\mathbf{H}_{\mathrm{R},k})$ as follows
\begin{eqnarray}
\label{decoupletrans}
\mathrm{Tr}(\bm{\Psi}\mathbf{G}\mathbf{W}_r \mathbf{G}^H\bm{\Psi}^H\mathbf{H}_{\mathrm{R},k})=&&\hspace*{-6mm}\frac{1}{2}\left \|\bm{\Psi}+\mathbf{G}\mathbf{W}_r \mathbf{G}^H\bm{\Psi}^H\mathbf{H}_{\mathrm{R},k}\right \|_F^2-\frac{1}{2}\mathrm{Tr}( \bm{\Psi}^H\bm{\Psi})\notag\\
    -&&\hspace*{-6mm}\frac{1}{2}\mathrm{Tr}(\mathbf{H}_{\mathrm{R},k}^H \bm{\Psi}\mathbf{G}\mathbf{W}_r^H\mathbf{G}^H   \mathbf{G}\mathbf{W}_r \mathbf{G}^H\bm{\Psi}^H\mathbf{H}_{\mathrm{R},k}).
\end{eqnarray}
We note that the right-hand side term of \eqref{decoupletrans} contains a bilinear function of optimization variables $\mathbf{W}_r$ and $\bm{\Psi}$, i.e., $\mathbf{G}\mathbf{W}_r \mathbf{G}^H\bm{\Psi}^H\mathbf{H}_{\mathrm{R},k}$, which is still non-convex. To circumvent this challenge, we further define a new optimization variable $\mathbf{Z}_r=\mathbf{W}_r \mathbf{G}^H\bm{\Psi}^H$, where $\mathbf{Z}_r\in\mathbb{C}^{N_\mathrm{T}\times M}$. Then, we introduce the following lemma to transform the constraint $\mathbf{Z}_r=\mathbf{W}_r \mathbf{G}^H\bm{\Psi}^H$ to a more tractable form.
\par
\textit{Lemma 1}:\hspace*{1mm}The equality constraint $\mathbf{Z}_r=\mathbf{W}_r \mathbf{G}^H\bm{\Psi}^H$ is equivalent to the following inequality constraints:
\begin{eqnarray}
&&\mbox{C5:}\hspace*{1mm}\begin{bmatrix}
\mathbf{U}_r & \mathbf{Z}_r  & \mathbf{W}_r \mathbf{G}^H \\ 
\mathbf{Z}_r^H & \mathbf{V}_r & \bm{\Psi}\\ 
\mathbf{G}\mathbf{W}_r^H & \bm{\Psi}^H & \mathbf{I}_M
\end{bmatrix}\succeq\mathbf{0},\hspace*{1mm}\forall r\in\mathcal{K},\\[-2mm]\notag\\
&&\mbox{C6:}\hspace*{1mm}\mathrm{Tr}\left(\mathbf{U}_r-\mathbf{W}_r \mathbf{G}^H\mathbf{G}\mathbf{W}_r^H\right)\leq 0,\hspace*{1mm}\forall r\in\mathcal{K},
\end{eqnarray}
where $\mathbf{U}_r\in\mathbb{C}^{N_\mathrm{T}\times N_\mathrm{T}}$ and $\mathbf{V}_r\in\mathbb{C}^{M\times M}$ are auxiliary optimization variables.
\par
\textit{Proof:~}The equality constraint $\mathbf{Z}_r=\mathbf{W}_r \mathbf{G}^H\bm{\Psi}^H$ has a similar structure as the constraint in \cite[Eq. (3)]{6698281} and Lemma 1 can be proved by closely following the same steps as in \cite[Appendix A]{6698281}. Due to the space limitation, we omit the detailed proof of Lemma 1.
\par
\subsection{Inner Approximation}
After employing the proposed bilinear transformation, we can rewrite the right-hand side term of \eqref{decoupletrans} as follows
\begin{eqnarray}
    \frac{1}{2}\left \|\bm{\Psi}+\mathbf{G}\mathbf{Z}_r\mathbf{H}_{\mathrm{R},k}\right \|_F^2-\frac{1}{2}\mathrm{Tr}\left( \bm{\Psi}^H\bm{\Psi}\right)-\frac{1}{2}\mathrm{Tr}\left(\mathbf{H}_{\mathrm{R},k}^H \mathbf{Z}_r^H\mathbf{G}^H\mathbf{G}\mathbf{Z}_r\mathbf{H}_{\mathrm{R},k}\right).
\end{eqnarray}
We note that the quadratic terms $\mathrm{Tr}( \bm{\Psi}^H\bm{\Psi})$ and $\mathrm{Tr}(\mathbf{H}_{\mathrm{R},k}^H \mathbf{Z}_r^H\mathbf{G}^H\mathbf{G}\mathbf{Z}\mathbf{H}_{\mathrm{R},k})$ are obstacles for efficient algorithm design. To handle this issue, we construct respective global underestimators for these terms by employing their first-order Taylor approximations via the iterative IA approach. In particular, we have
\begin{eqnarray}
    &&\mathrm{Tr}\left (\bm{\Psi}^H\bm{\Psi}\right )\geq \mathrm{Tr}\left (\left (2\bm{\Psi}^{(j)}\right )^H\bm{\Psi}\right )-\left \|\bm{\Psi}^{(j)} \right \|_F^2,\label{taylorapproximation1}\\
    &&\mathrm{Tr}\left (\mathbf{H}_{\mathrm{R},k}^H \mathbf{Z}_r^H\mathbf{G}^H\mathbf{G}\mathbf{Z}_r\mathbf{H}_{\mathrm{R},k}\right )\geq\mathrm{Tr}\left (\left (2\mathbf{H}_{\mathrm{R},k}^H\mathbf{G}^H\mathbf{G}\mathbf{Z}_r^{(j)}\mathbf{H}_{\mathrm{R},k}\right )^H\mathbf{Z}_r\right)-\left \|\mathbf{G}\mathbf{Z}_r^{(j)}\mathbf{H}_{\mathrm{R},k}\right \|_F^2,\label{taylorapproximation2}
\end{eqnarray}
where $\bm{\Psi}^{(j)}$ and $\mathbf{Z}_r^{(j)}$ are intermediate solutions obtained in the $j$-th iteration and superscript $j$ denotes the iteration index of the optimization variables. Moreover, by applying steps similar to \eqref{vectornormtotrace}, \eqref{decoupletrans}, \eqref{taylorapproximation1}, and \eqref{taylorapproximation2}, we construct an upper bound for the term $-\left | (\mathbf{h}_{\mathrm{D},k}^H+\mathbf{h}_{\mathrm{R},k}^H\mathbf{A}\bm{\Theta}\mathbf{G})\mathbf{w}_k\right |^2$ in constraint C1. As a result, a convex subset of constraint C1 is obtained as

\begin{eqnarray}
    \overline{\mbox{C1}}\mbox{:}\hspace*{1mm}&&\frac{\Gamma_{\mathrm{req}_k}}{2}\underset{r\in\mathcal{K}\setminus \left \{k\right \}}{\sum }\left \|\bm{\Psi}+\mathbf{G}\mathbf{Z}_r\mathbf{H}_{\mathrm{R},k}\right \|_F^2-[\Gamma_{\mathrm{req}_k}(K-1)-1]\left [\mathrm{Tr}\left (\left (\bm{\Psi}^{(j)}\right )^H\bm{\Psi}\right )-\frac{1}{2}\left \|\bm{\Psi}^{(j)} \right \|_F^2\right ]
    \notag\\
    &&-\Gamma_{\mathrm{req}_k}\underset{r\in\mathcal{K}\setminus \left \{k\right \}}{\sum }\left [\mathrm{Tr}\left (\left (\mathbf{H}_{\mathrm{R},k}^H\mathbf{G}^H\mathbf{G}\mathbf{Z}_r^{(j)}\mathbf{H}_{\mathrm{R},k}\right )^H\mathbf{Z}_r\right)-\frac{1}{2}\left \|\mathbf{G}\mathbf{Z}_r^{(j)}\mathbf{H}_{\mathrm{R},k}\right \|_F^2\right ]\notag\\
    &&+\Gamma_{\mathrm{req}_k}\left (\underset{r\in\mathcal{K}\setminus \left \{k\right \}}{\sum }\mathrm{Tr}(\mathbf{H}_{\mathrm{D},k}\mathbf{W}_r)+\sigma_d^2\mathrm{Tr}(\bm{\Psi}^H\mathbf{H}_{\mathrm{R},k}\bm{\Psi} )+\sigma_{n_k}^2\right )-\mathrm{Tr}(\mathbf{H}_{\mathrm{D},k}\mathbf{W}_k)\notag\\
    &&-\frac{1}{2}\left \|\bm{\Psi}+\mathbf{G}\mathbf{Z}_k\mathbf{H}_{\mathrm{R},k}\right \|_F^2+\mathrm{Tr}\left (\left (\mathbf{H}_{\mathrm{R},k}^H\mathbf{G}^H\mathbf{G}\mathbf{Z}_k^{(j)}\mathbf{H}_{\mathrm{R},k}\right )^H\mathbf{Z}_k\right)-\frac{1}{2}\left \|\mathbf{G}\mathbf{Z}_k^{(j)}\mathbf{H}_{\mathrm{R},k}\right \|_F^2\notag\\
    &&+\mathrm{Tr}\left (
    \widetilde{\mathbf{h}}_k\widetilde{\mathbf{h}}_k^H
    \begin{bmatrix}
    \mathbf{0} & \Gamma_{\mathrm{req}_k}\underset{r\in\mathcal{K}\setminus \left \{k\right \}}{\sum }\mathbf{Z}_r^H-\mathbf{Z}_k^H \\ 
    \Gamma_{\mathrm{req}_k}\underset{r\in\mathcal{K}\setminus \left \{k\right \}}{\sum }\mathbf{Z}_r-\mathbf{Z}_k & \mathbf{0}
    \end{bmatrix}\right )\leq 0,\hspace*{2mm}\forall k,
\end{eqnarray}
where $\widetilde{\mathbf{h}}_k^H\in\mathbb{C}^{1\times (M+N_\mathrm{T})}$ is defined as $\widetilde{\mathbf{h}}_k^H=[\mathbf{h}_{\mathrm{R},k}^H  
\hspace*{2mm}\mathbf{h}_{\mathrm{D},k}^H]$. Similarly, constraint C2 can be approximated by the following convex constraint:
\begin{eqnarray}
    \overline{\mbox{C2}}\mbox{:}\hspace*{1mm}&&\underset{k\in\mathcal{K}}{\sum }\left[\frac{1}{2}\left \|\bm{\Psi}+\mathbf{G}\mathbf{Z}_k\right \|_F^2-\mathrm{Tr}\left (\left (\mathbf{G}^H\mathbf{G}\mathbf{Z}_k^{(j)}\right )^H\mathbf{Z}_k\right)+\frac{1}{2}\left \|\mathbf{G}\mathbf{Z}_k^{(j)}\right \|_F^2\right]\notag\\
    &&-K\left[\mathrm{Tr}\left (\left (\bm{\Psi}^{(j)}\right )^H\bm{\Psi}\right )-\frac{1}{2}\left \|\bm{\Psi}^{(j)} \right \|_F^2\right]+\sigma_d^2\mathrm{Tr}(\bm{\Psi}\bm{\Psi}^H)\leq P_{\mathrm{A}}.
\end{eqnarray}
\par
On the other hand, we note that constraint C6 is in the canonical form of a difference of convex functions which is a non-convex constraint. To tackle this obstacle, again, we construct a global underestimator of $\mathrm{Tr}(\mathbf{W}_r \mathbf{G}^H\mathbf{G}\mathbf{W}_r^H)$. Specifically, we have
\begin{equation}
    \mathrm{Tr}(\mathbf{W}_r \mathbf{G}^H\mathbf{G}\mathbf{W}_r^H)\geq-\left \|\mathbf{W}_r^{(j)} \mathbf{G}^H \right \|_F^2+2\mathrm{Tr}\left((\mathbf{G}^H\mathbf{G}\mathbf{W}_r^{(j)})^H\mathbf{W}_r\right).
\end{equation}
Then, constraint C6 can be approximated by the following convex constraint:
\begin{equation}
    \overline{\mbox{C6}}\mbox{:}\hspace*{1mm}\mathrm{Tr}\left(\mathbf{U}_r\right)+\left \|\mathbf{W}_r^{(j)} \mathbf{G}^H \right \|_F^2-2\mathrm{Tr}\left((\mathbf{G}^H\mathbf{G}\mathbf{W}_r^{(j)})^H\mathbf{W}_r\right)\leq 0,\hspace*{1mm}\forall r\in\mathcal{K}.
\end{equation}
\par
Therefore, the optimization problem to be solved in the $(j+1)$-th iteration of the IA-based algorithm is given by
\begin{eqnarray}
\label{prob3}
\hspace*{2mm}&&\underset{\substack{\bm{\Psi},\mathbf{W}_k\in\mathbb{H}^{N_{\mathrm{T}}},\\\mathbf{Z}_k,\mathbf{U}_k,\mathbf{V}_k}}{\mino} \,\, \,\, \hspace*{2mm}F(\mathbf{W}_k)\overset{\Delta }{=}\underset{k\in\mathcal{K}}{\sum }\mathrm{Tr}(\mathbf{W}_k)\notag\\
&&\mbox{subject to}\hspace*{6mm}
\overline{\mbox{C1}},\overline{\mbox{C2}},\mbox{C3},\mbox{C4},\mbox{C5},\overline{\mbox{C6}}.
\end{eqnarray}
We note that the only obstacle to efficiently solving \eqref{prob3} is the rank-one constraint C4. To convexify the optimization problem in \eqref{prob3}, we apply SDR and remove constraint C4 from the formulation. Then, the resulting relaxed version of \eqref{prob3} becomes a standard convex optimization problem which can be optimally solved by convex program solvers such as CVX \cite{grant2008cvx}. Next, we introduce the following theorem to reveal the tightness of SDR.
\par
\textit{Theorem 1:~}Given any positive $\Gamma_{\mathrm{req}_k}$, the optimal beamforming matrix obtained from \eqref{prob3}, i.e., $\mathbf{W}_k^*$, is always a rank-one matrix.
\par
\textit{Proof:~}Problem \eqref{prob3} has a similar structure as \cite [Problem (17)]{yu2020power} and Theorem 1 can be proved following the same steps as in \cite [Appendix]{yu2020power}. The detailed proof of Theorem 1 is omitted for brevity. \qed
\par
\begin{algorithm}[t]
\caption{IA-based Algorithm}
\begin{algorithmic}[1]
\small
\STATE Set initial point $\mathbf{W}_k^{(j)}$, $\bm{\Psi}^{(j)}$, $\mathbf{Z}_k^{(j)}$, $\mathbf{U}_k^{(j)}$, $\mathbf{V}_k^{(j)}$, iteration index $j=1$, and error tolerance $0<\epsilon\ll1$.
\REPEAT
\STATE For given $\mathbf{W}_k^{(j)}$, $\bm{\Psi}^{(j)}$, $\mathbf{Z}_k^{(j)}$, $\mathbf{U}_k^{(j)}$, $\mathbf{V}_k^{(j)}$, obtain the intermediate solution $\mathbf{W}_k^{(j+1)}$, $\bm{\Psi}^{(j+1)}$, $\mathbf{Z}_k^{(j+1)}$, $\mathbf{U}_k^{(j+1)}$, $\mathbf{V}_k^{(j+1)}$ by solving the rank constraint-relaxed version of problem \eqref{prob3}
\STATE Set $j=j+1$
\UNTIL $\frac{F(\mathbf{W}_k^{(j-1)})-F(\mathbf{W}_k^{(j)})}{F(\mathbf{W}_k^{(j)})}\leq \epsilon$
\end{algorithmic}
\end{algorithm}
\par
We summarize the proposed algorithm in \textbf{Algorithm 1}. Note that the objective function of \eqref{prob3} is monotonically non-increasing in each iteration of \textbf{Algorithm 1}. Moreover, according to \cite[Theorem 1]{marks1978general}, the proposed algorithm is guaranteed to converge to a locally optimal solution of \eqref{prob1} in polynomial time. The per iteration computational complexity of \textbf{Algorithm 1} is given by $\mathcal{O}\Big(\mathrm{log}(1/\epsilon)\big((3K+1)^3+(3K+1)^2N_{\mathrm{T}}^2+(3K+1)N_{\mathrm{T}}^3+(2K+1)^3+(2K+1)^2M^2+(2K+1)M^3\big)\Big)$, where $\mathcal{O}\left ( \cdot  \right )$ is the big-O notation \cite[Theorem 3.12]{polik2010interior} and $\epsilon$ is the convergence tolerance of \textbf{Algorithm 1}. 
\section{Simulation Results}
\begin{table}[t]\vspace*{0mm}\caption{System simulation parameters.}\vspace*{-2mm}\label{tab:parameters}\footnotesize
\newcommand{\tabincell}[2]{\begin{tabular}{@{}#1@{}}#2\end{tabular}}
\centering
\begin{tabular}{|l|l|l|}\hline
    \hspace*{-1mm}$f_c$ & Carrier center frequency & $2.4$ GHz \\
\hline
    \hspace*{-1mm}$\sigma_k^2$& Noise power at the users & $-114$ dBm \\
\hline
    \hspace*{-1mm}$\sigma_d^2$& Dynamic noise power & $-100$ dBm \cite{zhang2021active}\\
\hline
    \hspace*{-1mm}$\epsilon$ & Convergence tolerance & $10^{-3}$ \\
\hline
\end{tabular}
\vspace*{-2mm}
\end{table}
In this section, the system performance of the
proposed resource allocation scheme is evaluated via simulations. 
The BS is equipped with $N_{\mathrm{T}}=4$ antennas and serves one sector of a cell with a radius of $R$ m, where $K=3$ users are randomly and uniformly distributed in this sector. The active IRS comprises $M$ elements and is deployed at the edge of the sector. Moreover, the fading coefficients of all the channels are generated as independent and identically distributed Rician random variables with Rician factor $3$ dB. In addition, the path loss exponents for the direct links and the reflected links between the BS and the users are $\alpha_{\mathrm{d}}$ and $\alpha_{\mathrm{r}}$, respectively. For ease of presentation, we assume that the minimum required SINRs of all users are identical, i.e., $\Gamma_{\mathrm{req}_k}=\Gamma_{\mathrm{req}}$, $\forall k$. The adopted simulation parameter values are listed in Table \ref{tab:parameters}.
\par
For comparison, we consider two baseline schemes. For baseline scheme 1, we assume that an IRS is not deployed. Then, we optimize the beamforming vector $\mathbf{w}_k$ for minimization of the transmit power at the BS. For baseline scheme 2, we divide the power available at the active IRS, $P_{\mathrm{A}}$, equally among the IRS elements, i.e., $a_m=\sqrt{\frac{P_{\mathrm{A}}}{M}}$, $\forall m\in\mathcal{M}$, and generate the phases of the IRS elements in a random manner. Moreover, we adopt zero-forcing (ZF) beamforming at the BS. Then, we solve a problem similar to problem \eqref{prob1}, where we optimize the power allocated to user $k$, i.e., $p_k\in\mathbb{R}_+$.
\subsection{Transmit Power Minimization}
\begin{figure}[t]\vspace*{-2mm}
\centering
\includegraphics[width=3.8in]{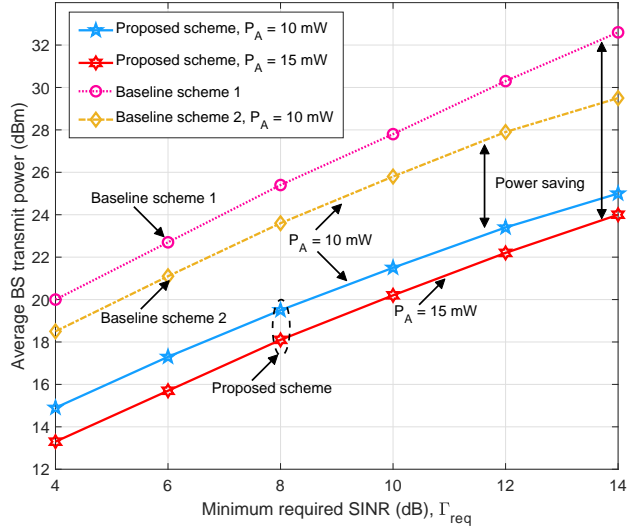}
\vspace*{-4mm}
\caption{Average BS transmit power (dBm) versus minimum required SINR of the users for $K=3$, $N_{\mathrm{T}}=4$, $M=10$, $\alpha_{\mathrm{d}}=3.8$, $\alpha_{\mathrm{r}}=2.3$, and $R=100$ m.}\vspace*{-2mm}\label{powersinr_ActiveIRS}
\end{figure}
\par
In Figure \ref{powersinr_ActiveIRS}, we investigate the average BS transmit power versus the minimum required SINR of the users for a scenario where the direct links are severely shadowed ($\alpha_{\mathrm{d}}=3.8$). We can observe from Figure \ref{powersinr_ActiveIRS} that the transmit power of the proposed scheme and the two baseline schemes monotonically increases with $\Gamma_{\mathrm{req}}$. This is attributed to the fact that to satisfy a more stringent minimum SINR requirement, the BS has to transmit with a higher power. Yet, the proposed scheme yields substantial power savings compared to the two baseline schemes even if we account for the total transmit power. For example, for $\Gamma_{\mathrm{req}}=4$, the proposed scheme with $P_{\mathrm{A}}=10$ mW consumes $10^{(1.5)}+10\approx41.6$ mW, while baseline scheme 1 and baseline scheme 2 require $100$ mW and $73.1$ mW, respectively. In particular, for baseline scheme 1, since there is no IRS, there are no degrees of freedom (DoFs) available for customizing favorable wireless channels. As for baseline scheme 2, both the BS and the active IRS cannot fully exploit the DoFs available for resource allocation due to the partially fixed beamforming policy and the randomly generated IRS phase shifts, respectively. This highlights the effectiveness of the proposed scheme for jointly optimizing the beamformers at the BS and the active IRS elements. Moreover, as expected, increasing the maximum power allowance at the active IRS from $10$ mW to $15$ mW leads to further transmit power savings at the BS. This is because the additional power budget at the active IRS can be utilized to facilitate more accurate beamforming and to mitigate multiuser interference in a more effective manner.
\par
\subsection{Energy Efficiency Evaluation}
\begin{figure}[t]\vspace*{-2mm}
 \centering
\includegraphics[width=3.8in]{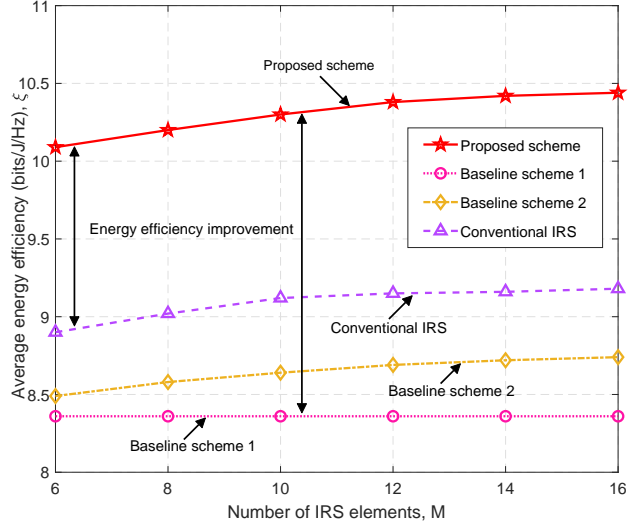}
\vspace*{-4mm}
\caption{Average energy efficiency versus the number of IRS elements with $K=3$, $N_{\mathrm{T}}=4$, $P_{\mathrm{A}}=20$ mW, $\Gamma_{\mathrm{req}}=10$ dB, $\alpha_{\mathrm{d}}=2.9$, $\alpha_{\mathrm{r}}=2.3$, and $R=200$ m.}\vspace*{-2mm}\label{powerSINR_SWIPT}
\end{figure}
To further investigate the performance of active IRSs, we also compare with a conventional IRS where the IRS elements just passively reflect the incident signals without amplification. In particular, we employ the IA-based algorithm developed in \cite{yu2020power} and solve a problem similar to \eqref{prob1} but replacing constraint C2 with a unit-modulus constraint induced by the passive IRS. For a fair comparison, we adopt the energy efficiency (bits/J/Hz) as the performance metric which is defined as\footnote{We set $P_\mathrm{A}=0$ when computing the energy efficiency of the system with the conventional passive IRS.} \cite[Eq. (19)]{yu2020power}
\begin{equation}
    \xi=\frac{\underset{k\in\mathcal{K}}{\sum }\mathrm{log}_{\mathrm{2}}(1+\Gamma_k)}{\frac{1}{\eta }\underset{k\in\mathcal{K}}{\sum }\left \|\mathbf{w}_k\right \|^2+N_\mathrm{T}P_\mathrm{T}+P_\mathrm{C}+MP_\mathrm{I}+\frac{1}{\eta }P_\mathrm{A}},
\end{equation}
where $\eta=0.5$ is the power amplifier efficiency, $P_\mathrm{T}=100$ mW is the circuit power that maintains one BS antenna element operational, $P_\mathrm{C}=85$ mW is the static circuit power of the BS \cite{yu2020power}, $P_\mathrm{I}=2$ mW is the circuit power required to support one IRS element\footnote{In this paper, we adopt the same $P_\mathrm{I}$ for passive and active IRS elements. Yet, in practice, depending on the specific hardware structure and components, active IRS elements may consume slightly more power for supporting the required amplifier \cite{lonvcar2019ultrathin}.} \cite{pei2021prototype}, and $P_\mathrm{A}=20$ mW is the power allowance of the active IRS \cite{zhang2021active}. Figure \ref{powerSINR_SWIPT} illustrates the average energy efficiency versus the number of IRS elements for a scenario where the direct links are slightly shadowed ($\alpha_{\mathrm{d}}=2.9$). As can be seen from Figure \ref{powerSINR_SWIPT}, the energy efficiencies of the proposed scheme, the scheme employing a conventional IRS, and baseline scheme 2 monotonically increase with the number of IRS elements. In particular, due to the low-power consumption of IRS phase shifters, deploying more IRS elements does not significantly increase the operational power of the IRS. Moreover, additional IRS elements introduce extra DoFs that can be exploited to proactively configure the wireless channel which yields transmit power savings. Besides, for the proposed scheme, additional IRS elements allow the active IRS to strike a balance between effectively mitigating the dynamic noise amplification and amplifying the desired signals. On the other hand, we observe that the proposed scheme outperforms the scheme employing a conventional passive IRS and the two baseline schemes. In particular, for the scenario where the direct links are slightly shadowed, deploying passive IRSs can not effectively enhance performance due to the double path loss effect. In contrast, the proposed scheme employing the active IRS can simultaneously adjust the phase and the amplitude of the reflected signal to combat the double path loss effect, which yields a performance enhancement at the expense of supplying extra power to the IRS. This observation strongly motivates the application of active IRSs to further improve the system performance, especially when the direct links are not weak. 
\section{Conclusion}
In this paper, we investigated the deployment of active IRSs, where, unlike conventional passive IRSs, each IRS element is equipped with an amplifier, and studied the resulting resource allocation algorithm design problem for a multiuser communication system. In particular, we jointly optimized the beamforming vectors at the BS and the IRS parameters for minimization of the BS transmit power. To tackle the formulated non-convex optimization problem, we developed a novel low-complexity algorithm, based on the bilinear transformation and IA. The developed algorithm is guaranteed to converge to a locally optimal solution of the considered problem. Simulation results showed that the proposed scheme achieves considerable power savings compared to two baseline schemes. Moreover, our results revealed that active IRSs are a promising means to combat the performance degradation caused by the double path loss effect in IRS-assisted communication systems. 
\vspace*{-1mm}
\bibliographystyle{IEEEtran}
\bibliography{Reference_List}
\end{document}